\documentstyle[epsf]{mn}
\def\epsfsize#1#2{\hsize}

\begin{document}
                                                                                
\title[Bolometric Corrections and Temperature scales for Population II 
giants]
{\centering Towards the absolute planes: a new calibration of the \\
Bolometric Corrections and Temperature scales \\for Population II Giants
\thanks{Based on data taken at the ESO--MPI 2.2m Telescope equipped 
with the near IR camera IRAC2 - ESO, La Silla (Chile).}}

\author[P. Montegriffo et al.]
       {P. Montegriffo,$^1$
       F.R. Ferraro,$^1$
       L. Origlia,$^1$
       and F. Fusi Pecci$^2$\thanks{on leave from Osservatorio Astronomico
       di Bologna}\\
        $^1$Osservatorio Astronomico di Bologna, Via Zamboni 33,
        40126 Bologna, ITALY\\
        $^2$Stazione Astronomica, 09012 Capoterra, Cagliari, ITALY}

\maketitle

\date{Accepted
      Received ;
      in original form }
\pubyear{1997}

\begin{abstract}
We present new determinations of bolometric corrections and effective 
temperature scales as a function of infrared  and optical colors, 
using a large database of photometric observations of about
6500 Population II giants in Galactic Globular Clusters (GGCs), 
covering a wide range in metallicity (--2.0$<$[Fe/H]$<$0.0). 
\par\noindent
New relations for 
BC$_K$ {\it vs} (V--K), (J--K) and  
BC$_V$ {\it vs} (B--V), (V--I), (V--J), and new calibrations for 
T$_{eff}$, using both an empirical relation and model atmospheres, 
are provided. 
\par\noindent
Moreover, an empirical relation to derive the R parameter of the 
Infrared Flux Method as a function of the stellar temperature is also
presented.
\end{abstract}


\section{Introduction}

A global test of stellar evolutionary models requires a direct comparison 
between theoretical tracks and observations for stars spanning a wide range 
in 
stellar parameters, such as temperature, luminosity and metallicity.
In order to achieve these goals at least two fundamental ingredients are 
needed:
\par
{\it i)~} a complete and homogeneous database of photometric observations;
\par
{\it ii)~} a suitable set of transformations between observables and 
absolute quantities.  

GGCs are the best empirical laboratory to obtain complete and homogeneous 
spectrophotometric information on Pop. II stars over a wide range of 
metallicities.

Rewieving the published works on 
the transformations to the absolute plane (see Sect.3), that is bolometric 
corrections (BCs) and temperature scales as a function of different colors,  
it is easy to see that very often these calibrations are not based on 
a complete and homogeneous set of data, spanning a wide range of stellar 
parameters. Moreover, {\it ad hoc} correction factors are usually adopted 
to take into account for example possible systematic differences between 
different photometric systems and/or different assumptions for the reference 
solar quantities, for the adopted model atmospheres, and for different 
laws to extrapolate the data towards the UV/IR ranges.

Such a scenario indicates that any calibration in the absolute
plane can hardly be fully self--consistent as it always depends
on the adopted transformations and, more crucial,  the residuals  among 
different scales are very rarely  linear with the involved parameters.

In order to improve the available determinations of the bolometric 
corrections 
and temperature scales, we use here our IR photometric database 
on GGC stars combined with available optical data from the 
literature to calibrate new, independent, (hopefully) self--consistent 
transformations, particularly useful to study the red stellar sequences 
in Pop II stars.

In Sect.2 we present the complete database used in our analysis 
which includes about 6500 RGB and HB stars in a sample of 10 GGCs 
observed in both optical and near IR bands.
In Sect.3 we derive the transformations from observed magnitudes and colors 
to absolute quantities, such as bolometric corrections and effective 
temperatures. All the results are listed in Table 3.
In Sect.4 we compare the inferred scales with 
existent ones and we give a fully empirical calibration of the R parameter 
of the Infrared Flux Method (IRFM) as a function of the effective 
temperature. 
Schematic conclusions are eventually presented in Sect.5.

\section{The database}

The IR database used in this study includes photometric observations in J 
and K bands of about 17000 stars belonging to 10 GGCs spanning the whole 
range in metallicity (from [Fe/H]=--2.15 to [Fe/H]$\approx $0.0). 
The data were obtained at ESO, La Silla (Chile), during two different runs 
(on June 1992 and June 1993), using the ESO--MPI 2.2m telescope and the 
near--IR camera IRAC--2 (Moorwood et al. 1992) equipped with a NICMOS--3 
256x256 array detector.
The complete description of the observations and data reduction 
can be found in Ferraro et al. (1994a,b) and Montegriffo et al. (1995) 
and in a series of forthcoming papers, where each individual cluster 
is discussed in more detail. 

Our IR database was cross-correlated with other optical and IR catalogs in 
the literature in order to provide complete UBVRIJK photometry of as many 
stars as possible. The final sample used in the following analysis 
includes about 6500  RGB and HB stars.
  
The reddening correction in each photometric band was performed starting
from the E(B--V) color excess (as reported by Armandroff (1989) with the
exception of  M68 and M69 for which we used the values by Walker 1994 
and Ferraro et al. 1994a, respectively) and using the extinction law 
proposed by Rieke \& Lebovsky (1985).  
In Table 1 the main features of the database are listed:
the GGC names, the cluster metallicity as quoted by Zinn (1985), 
the available photometric bands, and the adopted  E(B--V) color excess.
In the following the observed colors are always corrected for reddening. 
 
\subsection{Notes on individual clusters}

{\bf M15:~}UBVR photometry from Stetson (1994).
\vskip 0.1cm
\noindent   
{\bf M30:~}UBV photometry from Bergbusch (1996).   
\vskip 0.1cm
\noindent   
{\bf M68:~}BVI photometry from Walker (1994).
\vskip 0.1cm 
\noindent   
{\bf M55:~}BV photometry from Piotto (1996) and VI from Ortolani \& Desidera 
(1996): V was averaged.  
\vskip 0.1cm
\noindent   
{\bf M4:~}BV photometry from  Lee (1977).
\vskip 0.1cm 
\noindent   
{\bf M107:~}BV photometry from Ferraro et al. (1991).
\vskip 0.1cm
\noindent   
{\bf M69:~}BV photometry from  Ferraro et al. (1994a).
\vskip 0.1cm
\noindent   
{\bf 47Tuc:~}UBV photometry from Auriere \& Lauzeral (1996) and VI from  
Ortolani \& Desidera (1996). The VI photometry was compared with that one   
obtained by Da Costa \& Armandroff (1990) for the few stars in common. 
There is an excellent  agreement in V, while a systematic shift in the zero 
point by about 0.12 mag in I has been found: we applied this correction 
to the I photometry from  Ortolani \& Desidera (1996).
\vskip 0.1cm
\noindent   
{\bf NGC6553:~}VI photometry from  Guarnieri et al. (1997) using HST data. 
\vskip 0.1cm
\noindent   
{\bf NGC6528:~}VI photometry from  Ortolani et al. (1995) using HST data.

\section{Absolute quantities}

\begin{table}
 \centering
 \begin{minipage}{80mm} 
  \caption{ Adopted  IR and optical database.}
  \begin{tabular}{@{}lcccc@{}}
Cluster & & [Fe/H]$^a$ & Photometry$^b$ & E(B--V)$^c$\\ 
NGC7078 & M15    & --2.15 &  UBVRJK   &0.10 \\  
NGC7099 & M30    & --2.13 &  UBVJK    &0.04 \\  
NGC4590 & M68    & --2.09 &  BVIJK    &0.07 \\  
NGC6809 & M55    & --1.82 &  BVIJK    &0.06 \\  
NGC6121 & M4     & --1.19 &  BVJK    &0.40 \\  
NGC6171 & M107   & --0.89 &  BVJK    &0.31 \\  
NGC6637 & M69    & --0.75 &  BVJK    &0.17 \\  
NGC104  & 47Tuc  & --0.70 &  UBVIJK   &0.04 \\  
NGC6553 &        & --0.29 &  VIJK     &0.78 \\  
NGC6528 &        & --0.07 &  VIJK     &0.56 \\  
\end{tabular}
\vspace {0.2cm}
{\footnotesize
\par\noindent
{\bf $^a$~}From Zinn (1985).   
\par\noindent
{\bf $^b$~}J,K from our IR--array survey, U,B,V,R,I from the literature 
(refers to  Sect.2.1 for the bibliographic sources).
\par\noindent
{\bf $^c$~}E(B--V) from Armandroff (1989), but M68 from Walker (1994) 
and M69 from Ferraro et al. (1994a), respectively.
}
\end{minipage}
\end{table}

\begin{table*}
 \centering
 \begin{minipage}{105mm}
  \caption{ Absolute flux calibrations of a zero magnitude star in UBVRIJHKL
            bands for different photometric systems. Wavelengths are
            in $\mu$m and fluxes in units of 10$^{-8}$ erg s$^{-1}$ 
            cm$^{-2}$ $\mu $m$^{-1}$.} 
  \begin{tabular}{@{}lcccccc@{}}
Filter  &\hskip 25pt Johnson$^a$&  &\hskip 20pt Bessell $^b$&&\hskip 20pt 
ESO$^c$ &\\ 
&\hskip -0pt $\lambda _{eff}$&\hskip -5pt f$_{\lambda _0}$& 
\hskip -3pt $\lambda _{eff}$& \hskip -5pt f$_{\lambda _0}$&
$\hskip -3pt \lambda _{eff}$&\hskip -5pt f$_{\lambda _0}$\\  
U &  0.36&  4345 &	 0.366&   4175 &     0.363&  4080 \\   
B &  0.44&  7194 &	 0.438&   6320 &     0.438&  6490 \\    
V &  0.55&  3917 &	 0.545&   3631 &     0.548&  3650 \\   
R &  0.70&  1762 &	 0.641&   2177 &     0.641&  2190 \\   
I &  0.90&  829.9 &	 0.798&   1126 &     0.795&  1180 \\ 
J &  1.25&  289.7 &	 1.221&	 314.7 &     1.244&  312 \\    
H &  1.62&  107.9 &	 1.632&	 113.8 &     1.634&  120 \\   
K &  2.20&  38.02 &	 2.187&	 39.61 &     2.190&  41.7 \\   
L &  3.50&  7.834 &	 3.451&	 7.080 &     3.770&  5.41 \\  
\end{tabular}
\vspace {0.2cm}
{\footnotesize
\par\noindent
{\bf $^a$~}Johnson system revised by Buzzoni (1996).
\par\noindent
{\bf $^b$~}Bessell system adopted in BCP97 (Castelli 1997).   
\par\noindent
{\bf $^c$~}UBVRI from Bessell (1990) and from Megessier (1995).                         
}
\end{minipage}
\end{table*}

Various calibrations of the bolometric correction and temperature scale have 
been proposed by different authors since many years. 
 
Johnson (1966) used a sample of 15 giants observed in the UBVRIJKLMN to 
get bolometric corrections in the V band, integrating the spectral energy 
distribution and applying small corrections ($<$0.1 mag) in the wavelength 
ranges not covered by his survey (such as for example the H band) to take 
into account possible molecular blending in cool stars. 
The zero point is defined as BC$^{\odot}_V$=0.00.

Carney \& Aaronson (1979) reconstructed the energy distribution of a sample 
of dwarfs and subdwarfs by means of polynomial fits to the optical 
photometry, while the UV range was 
extrapolated using the Kurucz (1979) models  and the IR region
using  blackbodies. Their zero point is BC$^{\odot}_V$=--0.12.

Frogel, Persson \& Cohen (1981, hereafter FPC81) performed trapezoidal 
integrations of the energy distribution of a sample of stars observed in 
the UBVRIJHKL, and extrapolating the UV and IR fluxes. 
They adopted as zero point the value BC$^{\odot}_V$=--0.08.

More recently, Tinney et al. (1993) reconstructed the spectral energy 
distribution of cool stars combining photometric data in the VIJHKLL' bands 
with low resolution spectra. 

Alonso et al. (1995) determined 
the bolometric corrections in the K band of a sample of F,G,K dwarfs. 
They performed trapezoidal integrations using UBVRIJHK photometric data 
and included a correction factor C = f (T$_{eff}$, log~g, [Fe/H]) 
to the UV and IR fluxes as a function of the stellar parameters,  
using the Kurucz (1993) models.

\medskip
Concerning the determination of the stellar effective temperature, the only 
pure experimental way to derive it is to know the star intrinsic luminosity 
and apparent angular diameter. 
Ridgway et al. (1979) for example measured the angular diameter of a sample 
of nearby Pop I giants by means of lunar 
occultations, while Di Benedetto \& Rabbia (1987, hereafter DBR87) 
and Di Benedetto (1993, hereafter DB93) used interferometric techniques. 

Once the apparent stellar diameters have been measured, the effective 
stellar temperatures can be estimated assuming a proper scale of bolometric 
corrections. 
DBR87 have shown how an error of 0.05 mag in the 
determination of the BC$_K$ transfers into an error of $\approx$60 K 
in the temperature estimates for T$_{eff}$ around 5000 K and 
of $\approx$35 K for T$_{eff}\approx$3000 K. These uncertainties are 
of the same order of magnitude as the empirical errors, hence an accurate 
estimate of BC$_K$ is an important issue and severely affects the 
overall accuracy of the temperature estimated using this technique.

Another method which allows one to derive stellar temperatures in more 
distant objects is represented by the so--called IRFM 
proposed by Blackwell, Shallis \& Selby (1979) and Blackwell 
\& Lynas--Gray (1994). 

This method uses as a fundamental parameter the quantity 
$R~=f_{bol}/f_{\lambda }$, where $f_{bol}$ is the observed bolometric flux 
and $f_{\lambda}$ is the observed flux in a certain photometric band.   
Such a quantity is compared with the corresponding theoretical one, 
derived by means of 
models of atmospheres, in order to derive a temperature scale. 
Tinney et al. (1993) for their sample of dwarfs (see above) derived 
T$_{eff}$ by subsequent iterations of the IRFM.

Finally, a  fully theoretical approach is the extensive use of model
atmospheres to derive synthetic colors for different input stellar 
parameters.

Before describing the adopted procedures to calibrate our relations, we want 
to add a further comment on a crucial point: each absolute flux calibration 
requires a reference zero--magnitude star in all the explored photometric 
bands, that is in the specific photometric system used. 
As pointed out by many authors (see e.g. Bessell \& Brett 1988; 
Buzzoni 1989; 
Bessell 1990, Megessier 1995) the photometric CCD+filter characterization is 
often provided with a quite  large uncertainty (typically $\approx$5\%), 
and 
this may severely affect the accuracy of the derived quantities, particularly 
the bolometric fluxes, based on this instrumental calibration.  
Moreover, it is difficult to reconstruct the adopted values from 
the various authors since very often they are not tabulated. 
Since in the calibrations of the present paper we 
worked with different photometric systems, in Table 2 we 
report all the adopted zero magnitude fluxes.

\subsection{The determination of bolometric corrections}

\subsubsection{The BC$_K$ vs (V--K) and (J--K)}

For all the observed stars in the selected sample of GGCs 
we computed the bolometric correction BC$_K$ in the K band by means 
of the relation: 
$$BC_K=m_{bol}-K$$
where m$_{bol}$ is the apparent bolometric magnitude.

m$_{bol}$ was computed as follows: 
first we convert the apparent magnitudes corrected for reddening   
into fluxes using the classical formula:
$$f_{\lambda }~=~f_{\lambda _0}~\cdot 10^{-0.4\cdot m_{\lambda }}$$
where $f_{\lambda _0}~$ is the absolute flux in a certain photometric band 
corresponding to the zero magnitude. We used the UBVJK filters            
of Johnson and RI of Cousins in the ESO standard photometric system.
In Table 2 we report the absolute flux calibrations of a zero 
magnitude star for all the observed photometric bands, for different 
photometric systems. The values we adopted in our analysis 
are those listed in columns \#6 and \#7.
  
The energy distribution was reconstructed starting from a set of Planck 
functions B$_{\lambda }$(T$_c$) computed between two adjacent observed 
bands, where T$_c$ 
is the corresponding color temperature. 
We want to stress that the color temperature is here simply a "working 
parameter" and it is not used to constrain the effective stellar 
temperature. 

The final dense grid was obtained interpolating between two adjacent 
bands 
and extrapolating in the UV and IR ranges not covered by the observations.  
The apparent bolometric magnitude was then computed by means of integration 
over the energy distribution as follows:
$$m_{bol}~=~-2.5~ log_{10} ~\int ~f_{\lambda } ~d\lambda  ~+~ ZP$$ 
The zero point (ZP) of the bolometric scale is given by
$$ZP~=~M_{bol}^{\odot }~+~2.5~log_{10}~(F_{bol}^{\odot })$$
so its actual value depends upon the assumed ${M^{\odot }}_{bol} $ and 
$F_{bol}^{\odot }$.
We used M$_{bol}^{\odot }=4.75$, M$^{\odot }_V$=4.82 
(so BC$_V^{\odot}$=--0.07),  
L$_{bol}^{\odot }=3.86 \times 10^{33}$ erg s$^{-1}$, as
reported by Armandroff (1989), and the relation: 
$F_{bol}^{\odot }=L_{bol}^{\odot }/(4\pi d^2)$, where d=10 pc.
We obtained ZP=--11.478.
\vskip 0.1 cm
The accuracy of the inferred bolometric magnitudes primarily depends on the 
number of observed bands since extrapolations over a wide wavelength range 
may represent a severe approximation of the true energy distribution. 

\begin{figure} 
\epsffile{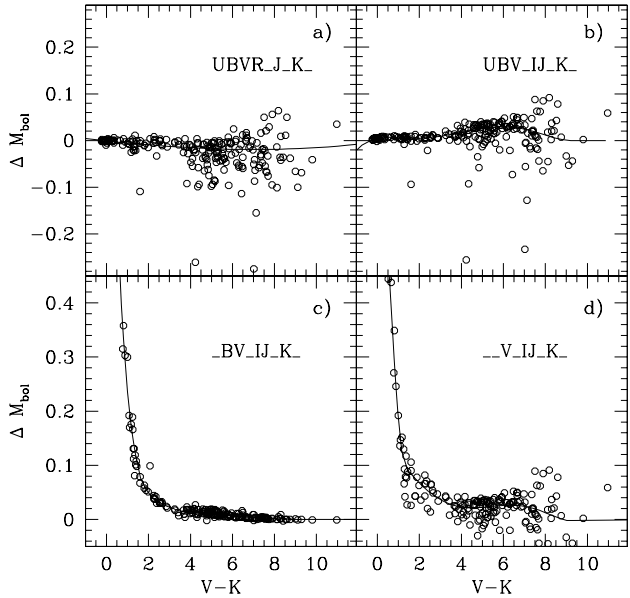}
\caption{
(a) $m_{bol}$ (tot) -- $m_{bol}$ (UBVRJK),
(b) $m_{bol}$ (tot)-- $m_{bol}$ (UBVIJK),
(c) $m_{bol}$ (tot)-- $m_{bol}$ (BVIJK), and  
(d) $m_{bol}$ (tot)-- $m_{bol}$ (VIJK)  {\it vs} 
the (V--K) color (see Sect.3.1.1). 
m$_{bol}$ (tot) was computed using the 
complete UBVRIJHKLMN set of filters. 
Open circles mark the stars by Morel \& Magnenat 
(1978), while the continuous lines are our best fits.}  
\end{figure}

In order to check the possible corrections to apply to the inferred values 
when a small number of bands was observed, we performed the following 
experiment: we used the catalog by Morel \& Magnenat (1978) which 
contains 212 
stars with a complete set of UBVRIJHKL photometry in the Johnson system 
and we applied our method to derive the apparent bolometric magnitudes with 
varying the number of available photometric bands, using the Johnson 
absolute flux calibration (columns \#2 and \#3 in Table 2). 

In Fig.1 we plot as an example a few simulations of the differences 
between $m_{bol}$ computed using all the Morel \& Magnenat (1978) bands  
and only those available in our GGC database {\it vs} the (V--K) color. 
Similar simulations were performed using (J--K).
The continuous line in the plots is our {\it numerical} best fit and 
represents 
the final corrections we applied to the bolometric magnitudes of the stars 
in our database as a function of their (V--K) and (J--K) colors and the 
available photometric bands.

On average, the discrepancies between the bolometric corrections computed 
using the entire set of photometric data and those computed using only 
a few bands are always within a few hundredths of a magnitude as one can see 
in Fig.1. Larger 
discrepancies, up to 0.1 mag, were found in the case of hot stars which 
lack U 
band observations: in this case, the UV continuum computed 
with a Planck function using the (B--V) color temperature is overestimated. 

\begin{figure} 
\epsffile{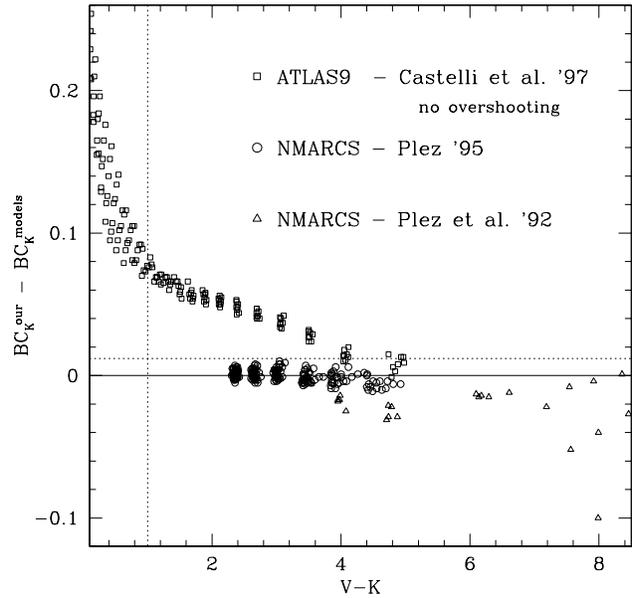}
\caption{
Differences between the inferred BC$_K$ using our procedure and that  
of the models, as a function of the (V--K) color. 
This comparison is performed using as reference database the set of 
synthetic 
grids of stellar atmospheres by BCP97. The dashed horizontal line is a 
systematic 
zero point shift of $\sim $0.01 between the two calibrations 
(our -- models) (see Sect.3.1.1).
At (V--K)$\le 1$ the Balmer discontinuity strongly affects the accuracy of
our BC$_K$.
}
\end{figure}

As a further check on the accuracy of the proposed method to derive 
the bolometric corrections we also reconstructed the energy distribution 
using direct trapezoidal integrations as suggested by many authors. 
The results of such a computation indicated that the corrections to apply 
if only a few colors are available should be even larger than in the case 
of integrations by means of Planck functions.

We did not apply any further correction due to the Balmer discontinuity 
in the UV and possible molecular sources of opacity in the IR.
Nevertheless, we performed some tests to evaluate the entity of such an 
effect using UBVRIJHKL synthetic color indices computed by Bessell, Castelli 
\& Plez (1997, hereafter BCP97) and derived from different grids of model 
atmospheres, which include both the ATLAS9 models computed without any 
overshooting for the convection (Castelli et al. 1997) 
and NMARCS models computed by Plez, Brett,
\& Nordlund (1992) and Plez (1995). 
The synthetic indices derived from Plez (1995) models
are computed for T$_{eff}$ ranging from 3600 K to 4750 K, gravity log g 
ranging from from --0.5 to 3.5,  and metallicities 
[Fe/H]=--0.6,--0.3,0.0,0.3, 0.6. 
Synthetic indices from ATLAS9 models and from 
Plez et al. (1992) models are tabulated only for the solar metallicity. 

\begin{figure} 
\epsffile{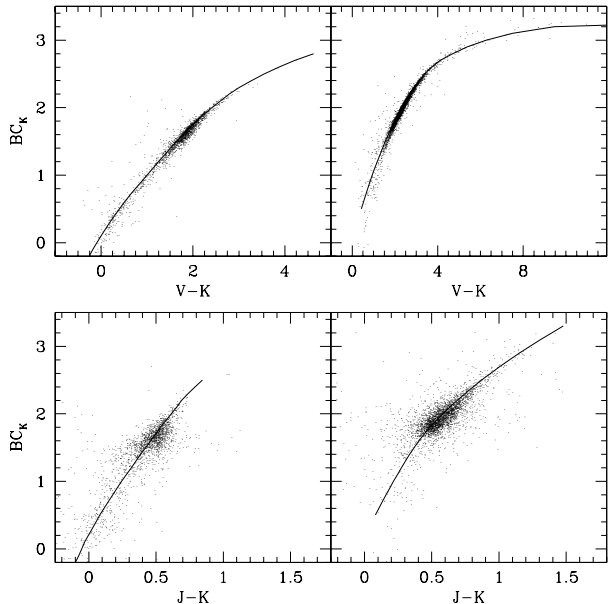}
\caption{
BC$_K$ as a function of (V--K) and (J--K) colors for metal poor 
(left panels) 
and metal rich (right panels) stars. 
Continuous lines are our best fit. 
}  
\end{figure}

We computed the BC$_K$ of these {\it synthetic } stars following the same 
procedure as for {\it true} stars in our database using the Bessell 
absolute flux calibration (columns \#4 and \#5 in Table 2). We compared 
the inferred values with those of the models themselves.
The results of this comparison are plotted in Fig.2 as a function 
of the (V--K) color. 
For (V--K)$>$1, the agreement between the two scales is satisfactory 
apart from an average shift in the zero point $\Delta $BC$_K$=0.01, 
possibly ascribed to the different assumption on M$_{bol}^{\odot }$ (they adopt 
M$_{bol}^{\odot }=4.74$).

For bluer colors the agreement between the two scales 
gets exponentially worse and worse since 
the continuum shape is strongly affected by the Balmer discontinuity 
and a Planck function is unable to properly reproduce it.

In order to investigate a possible dependence of the bolometric correction 
{\it vs} color relations on the metallicity, we divided the observed 
clusters into two groups: 
one at low metallicity ([Fe/H]$<$--1.0), using about 2000 stars in 
M15, M30, M68, 
M55 and M4, and the other at higher metallicity ([Fe/H]$>$ --1.0), 
using about 4500 stars in M107, M69, 47 Tuc, NGC6553 and NGC6528). 
We found a small (a few hundredths of a magnitude) 
difference when fitting our metal poor and metal rich stars in the 
empirical BC$_K$ {\it vs} (V--K) and (J--K) planes, more evident at 
(V--K)$<$2 and (J--K)$<$0.7.
This difference, that we want to stress is purely observational, 
could be explained in terms of a small metallicity dependence of 
the (V--K) and (J--K) colors as suggested by the model atmospheres whose 
continuum opacities, particularly at optical wavelengths, are slightly 
affected by changing the metal content.
A similar behaviour was found by Alonso et al. (1995) using an independent, 
semi--empirical approach to estimate the bolometric flux (see Sect.1). 

Nevertheless, since 
the difference we found is similar in size to the intrinsic photometric 
error, we cannot completely exclude the possibility that the observed effect 
is due to a systematic bias in the data .

In Fig.3, we plot the derived BC$_K$ 
as a function of the (V--K) and (J--K) colors 
for metal poor and metal rich stars separately.
Our best fits are shown as continuous lines.
The relation with the lowest spread is, as expected, that based on the 
(V--K) color which has the largest wavelength baseline.  

In Table 3 we report the corresponding BC$_K$ values as a function 
of the (V--K) and (J--K) colors from our best fits, both 
for metal poor and metal rich stars. The BC$_K$ is fixed (in other words 
we fixed the stellar temperature, see next section) and the colors slightly 
change ($\le$ 0.1 mag) in the two metallicity regimes.

\subsubsection {Other bolometric corrections}

\begin{figure} 
\epsffile{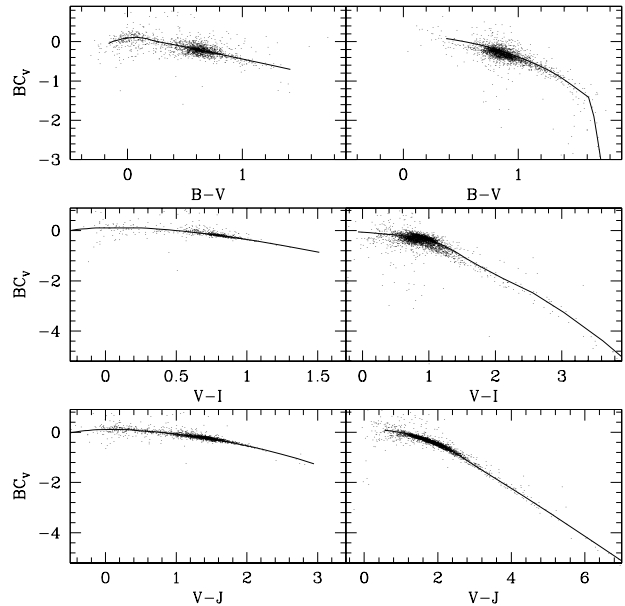}
\caption{
BC$_V$ vs (B--V), (V--I) and (V--J) colors for metal poor (left panels) 
and metal rich (right panels) stars.
Continuous lines are our best fits. 
}  
\end{figure}

Using our database we also inferred an estimate of BC$_V$ as a function 
of the 
(B--V), (V--I) and (V--J) colors, respectively, for metal poor 
and metal rich stars, as in the case of the BC$_K$.
In Fig.4, we plot the results and in Table 3 we listed the values for 
the corresponding BC$_K$.

The possible small dependence of the (V--K) color on metallicity should 
imply a similar dependence of the BC$_V$; hence, BC$_K$ could be a {\it purer} 
thermometer than BC$_V$.

Moreover, in the BC$_V$ {\it vs} (B--V) plane the difference between metal 
poor and metal rich stars is more pronounced (up to about 0.2 mag), 
since metallicity effects are more important at shorter wavelengths.     

\begin{table*}
 \centering
 \begin{minipage}{178mm} 
  \caption{BC$_K$ as a function of (V--K), (J--K), BC$_V$, (B--V), (V--I), 
(V--J) colors for metal poor and metal rich stars, respectively, and 
both empirical and theoretical T$_{eff}$.}
\end{minipage}
\begin{minipage}{260mm} 
%  \begin{tabular}{@{}ccccccccccccccccc@{}}
  \begin{tabular}{@{}rrrrrrrrrrrrrrrrr@{}}
 BC$_K$& ~~V--K&  J--K&  BC$_V$& B--V& V--I&  V--J&  ~~V--K& J--K&  BC$_V$ &  
B--V & V--I &  V--J &
~~T$_{eff}$ & T$_{eff}$\\
&{\underline {\bf Poor}}&&&&& &{\underline {\bf Rich}}&&&&&&~~Emp&Theor\\
 -0.40& ~-0.413& -0.137&  0.013& -0.118& -0.241& -0.420& ~--  &   --  &   --  &  
 --  &   --  &   --  & ~-- & 10462\\
 -0.30& ~-0.338& -0.119&  0.038& -0.089& -0.204& -0.360& ~--  &   --  &   --  &  
 --  &   --  &   --  & ~-- & 10139\\
 -0.20& ~-0.259& -0.099&  0.059& -0.062& -0.169& -0.300& ~--  &   --  &   --  &  
 --  &   --  &   --  & ~-- &  9824\\
 -0.10& ~-0.177& -0.075&  0.077& -0.040& -0.132& -0.237& ~--  &   --  &   --  &  
 --  &   --  &   --  & ~-- &  9518\\
  0.00& ~-0.091& -0.052&  0.091& -0.009& -0.098& -0.176& ~--  &   --  &   --  &  
 --  &   --  &   --  & ~-- &  9219\\
  0.10& ~-0.001& -0.030&  0.101&  0.033& -0.068& -0.120& ~--  &   --  &   --  &  
 --  &   --  &   --  & ~-- &  8928\\
  0.20& ~ 0.092& -0.003&  0.108&  0.068& -0.032& -0.064& ~--  &   --  &   --  &  
 --  &   --  &   --  & ~-- &  8645\\
  0.30& ~ 0.190&  0.026&  0.110&  0.078& -0.004&  0.157& ~--  &   --  &   --  &  
 --  &   --  &   --  & ~-- &  8370\\
  0.40& ~ 0.292&  0.055&  0.108&  0.078&  0.203&  0.234& ~--  &   --  &   --  &  
 --  &   --  &   --  & ~-- &  8102\\
  0.50& ~ 0.399&  0.082&  0.101&  0.100&  0.253&  0.319& ~0.425&  0.082&  0.075& 
  --  &   --  &   --  & ~7602&  7842\\
  0.60& ~ 0.510&  0.113&  0.090&  0.133&  0.293&  0.365& ~0.522&  0.107&  0.078& 
  --  &   --  &   --  & ~7386&  7589\\
  0.70& ~ 0.626&  0.145&  0.074&  0.168&  0.341&  0.436& ~0.621&  0.132&  0.079& 
 0.373&   --  &  0.563& ~7173&  7343\\
  0.80& ~ 0.745&  0.178&  0.055&  0.191&  0.390&  0.522& ~0.723&  0.159&  0.077& 
 0.378&   --  &  0.575& ~6963&  7105\\
  0.90& ~ 0.869&  0.210&  0.031&  0.218&  0.444&  0.615& ~0.829&  0.187&  0.071& 
 0.393&   --  &  0.611& ~6756&  6871\\
  1.00& ~ 0.996&  0.244&  0.004&  0.259&  0.498&  0.721& ~0.938&  0.216&  0.062& 
 0.413&   --  &  0.662& ~6553&  6642\\
  1.10& ~ 1.123&  0.279& -0.023&  0.307&  0.548&  0.828& ~1.050&  0.245&  0.050& 
 0.440&   --  &  0.724& ~6353&  6421\\
  1.20& ~ 1.249&  0.316& -0.049&  0.353&  0.592&  0.918& ~1.168&  0.274&  0.032& 
 0.476&   --  &  0.808& ~6156&  6208\\
  1.30& ~ 1.380&  0.350& -0.080&  0.406&  0.642&  1.018& ~1.289&  0.305&  0.011& 
 0.516&   --  &  0.897& ~5963&  6003\\
  1.40& ~ 1.506&  0.385& -0.106&  0.451&  0.681&  1.095& ~1.417&  0.336& -0.017& 
 0.564&   --  &  1.003& ~5774&  5806\\
  1.50& ~ 1.641&  0.423& -0.141&  0.510&  0.731&  1.193& ~1.549&  0.369& -0.049& 
 0.614& -0.067&  1.111& ~5589&  5618\\
  1.60& ~ 1.778&  0.461& -0.178&  0.572&  0.782&  1.289& ~1.688&  0.405& -0.088& 
 0.670&  0.111&  1.228& ~5407&  5438\\
  1.70& ~ 1.925&  0.498& -0.225&  0.651&  0.843&  1.402& ~1.834&  0.444& -0.134& 
 0.731&  0.321&  1.352& ~5229&  5264\\
  1.80& ~ 2.083&  0.532& -0.283&  0.747&  0.915&  1.531& ~1.987&  0.486& -0.187& 
 0.795&  0.561&  1.480& ~5056&  5097\\
  1.90& ~ 2.239&  0.572& -0.339&  0.838&  0.981&  1.647& ~2.147&  0.533& -0.247& 
 0.863&  0.761&  1.610& ~4887&  4933\\
  2.00& ~ 2.415&  0.612& -0.415&  0.961&  1.066&  1.792& ~2.317&  0.582& -0.317& 
 0.932&  0.900&  1.746& ~4723&  4774\\
  2.10& ~ 2.595&  0.653& -0.495&  1.088&  1.152&  1.934& ~2.498&  0.634& -0.398& 
 1.005&  1.027&  1.889& ~4563&  4618\\
  2.20& ~ 2.791&  0.690& -0.591&  1.239&  1.250&  2.090& ~2.691&  0.689& -0.491& 
 1.087&  1.133&  2.038& ~4406&  4468\\
  2.30& ~ 3.006&  0.738& -0.706&  1.416&  1.362&  2.263& ~2.900&  0.747& -0.600& 
 1.173&  1.226&  2.195& ~4254&  4321\\
  2.40& ~ 3.265&  0.790& -0.865&   --  &  1.512&  2.481& ~3.127&  0.807& -0.727& 
 1.263&  1.319&  2.359& ~4106&  4179\\
  2.50& ~ 3.551&  0.846& -1.051&   --  &   --  &  2.713& ~3.390&  0.870& -0.890& 
 1.362&  1.426&  2.544& ~3962&  4040\\
  2.60& ~ 3.860&   --  & -1.260&   --  &   --  &  2.950& ~3.703&  0.937& -1.103& 
 1.468&  1.559&  2.767& ~3819&  3904\\
  2.70& ~ 4.216&   --  & -1.516&   --  &   --  &   --  & ~4.117&  1.006& -1.417& 
 1.612&  1.762&  3.102& ~3667&  3771\\
  2.80& ~ 4.626&   --  & -1.826&   --  &   --  &   --  & ~4.696&  1.077& -1.896& 
 1.659&  2.111&  3.637& ~3523&  3639\\
  2.90& ~  --  &   --  &   --  &   --  &   --  &   --  & ~5.378&  1.152& -2.478& 
 1.690&  2.564&  4.268& ~3409&  3516\\
  3.00& ~  --  &   --  &   --  &   --  &   --  &   --  & ~6.263&  1.229& -3.263& 
 1.732&  3.033&  5.099& ~3317&  3385\\
  3.10& ~  --  &   --  &   --  &   --  &   --  &   --  & ~7.485&  1.310& -4.385& 
 1.793&  3.620&  6.258& ~3247&  3243\\
  3.20& ~  --  &   --  &   --  &   --  &   --  &   --  & ~9.444&  1.393& -6.244& 
  --  &  4.443&  8.142& ~3199&  2929\\
  3.30& ~  --  &   --  &   --  &   --  &   --  &   --  & ~19.746&  
1.478&-16.446&   --  &  8.119& 18.471&~3174&  2499\\
\end{tabular}
\end{minipage}
\end{table*}

\subsection{The determination of T$_{eff}$}

Our photometric database does not allow us to calibrate directly an empirical 
temperature scale since we do not know the angular 
diameter of the observed stars.
However DB93 provided a sample of field giants with measured 
angular diameters by means of Michelson interferometry and lunar 
occultations. We used his database, our BC  scale and zero point to 
calibrate a new empirical relation which links the 
effective temperature to the BC$_K$.

We also provided fully theoretical temperature scales as a function of 
suitable colors and BC, using the grids of stellar atmospheres computed by 
BCP97 at solar metallicity.

\subsubsection{Empirical T$_{eff}$}

We used the DB93's giants listed in his Table 1 and 3 of DB93, for which 
accurate estimates of the angular diameter exist, and by means 
of a second order polynomial fit we obtained V fluxes as a function 
of the (V--K) color, using our zero points. 
We did not use DB93's linear relations since they  
have a discontinuity around (V--K)=3.7 which propagates into the 
relation which gives T$_{eff}$ as a function of (V--K) and BC$_K$.

Following the procedure described in DB93 we derived the relation:

$ log~T_{eff}= 3.9619 - 0.0466 (V-K) + 0.0038  (V-K)^2 - $

$\hspace{16mm} 0.1  BC_K$

Since DB93's database is at solar metallicity the previous relation 
can be regarded as reliable only for metal rich stars.
Hence we computed the effective 
temperature 
using the latter relation only for the metal rich stars in our database  
and the results are reported in Fig.5 as 
a function of BC$_K$, while the mean relation is 
listed in Table 3. The average dispersion around the fiducial line 
is $\sim$30 K. Outside the 1$<BC_K<$3.1 range the relation was extrapolated.

\begin{figure} 
\epsffile{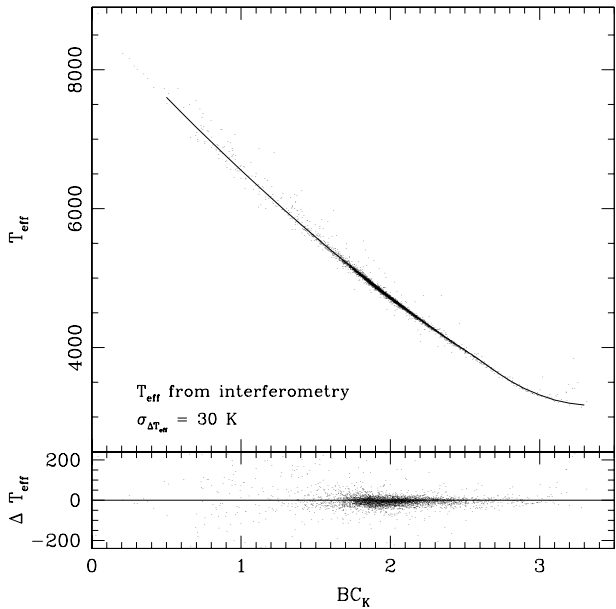}
\caption{
Empirical T$_{eff}$ as a function of BC$_K$ (upper panel) for metal rich 
stars. 
The continuous line is our best fit.
The dispersion (on average 30 K) 
around the fiducial line is also plotted (lower panel). 
}  
\end{figure}
 
\subsubsection{Theoretical T$_{eff}$}

The calibration of the effective temperature was performed using 
the grid of stellar atmospheres by BCP97 at solar metallicity. 
In order to evaluate the possible shift with varying metallicity we also 
used a grid of color indices for solar and 1/100 solar metallicity computed by 
Castelli (1997). These grids are based on the ATLAS9 model atmospheres in which 
the modification of the "approximate overshooting" to the mixing-length 
treatment of the convection was dropped. The indices slightly differ from the 
BCP97 indices owing to different passbands and zero points. 
For a given temperature and metallicity we interpolated the grids in gravity 
by means of the VandenBerg (1996) isochrones. 
For each star in our database we derived a mean temperature using both the 
(V--K), (V--J), (J--K) colors and the BC$_K$.

The relations were derived in the 
Bessell (1990) photometric system, as adopted by BCP97 to compute
their stellar atmosphere grids.
On the other hand, our observed database refers to the ESO photometric 
system. 
We used the relations quoted by Bessell \& Brett (1988) 
to transform the (V--K), (V--J) and (J--K) colors of the models from the 
Bessell system to the ESO one.
Moreover, since between the BC$_K$ computed with our procedure 
and those of the models 
a systematic shift in the zero point of 0.01 mag was found (see Sect.3.1.1 
and Fig.2), we added this shift to the BC$_K$ of the models. 
This avoids to have a systematic drift between 
T$_{eff}$ derived from the colors and the BC$_K$. 

In Fig.6 we plot the behaviour of the effective temperature as a function 
of both the (V--K), (V--J) and (J--K) colors in the ESO photometric system 
and the BC$_K$.
Our best fits (continuous lines) are plotted as well.
The spread visible in the T$_{eff}$ {\it vs} color distributions at the 
lowest temperature is due to a gravity effect. 
It demonstrates how small variations of 
such a parameter may strongly affect the accuracy of the derived 
temperatures below 3500 K.
This effect is less pronounced in the T$_{eff}$ 
{\it vs} BC$_K$ distribution. The latter is also the most linear 
and sensitive relation, particularly at low temperatures, 
confirming that BC$_K$ is probably the 
{\it purest stellar thermometer} for cool stars.

\begin{figure} 
\epsffile{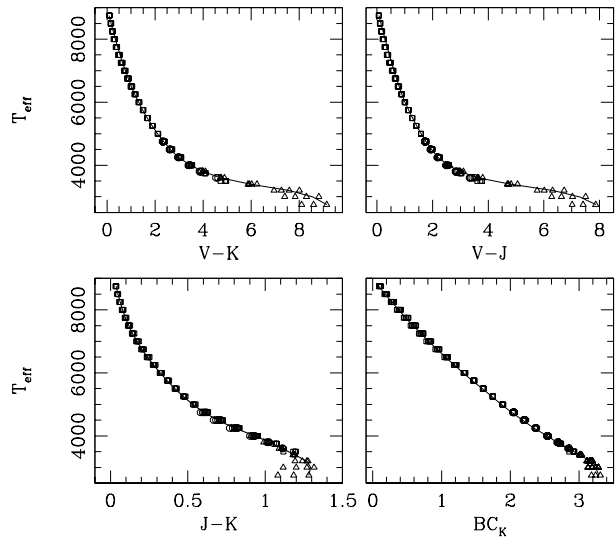}
\caption{
Theoretical T$_{eff}$ as a function of (V--K), (V--J), (J--K) and BC$_K$.
}  
\end{figure}

\begin{figure} 
\epsffile{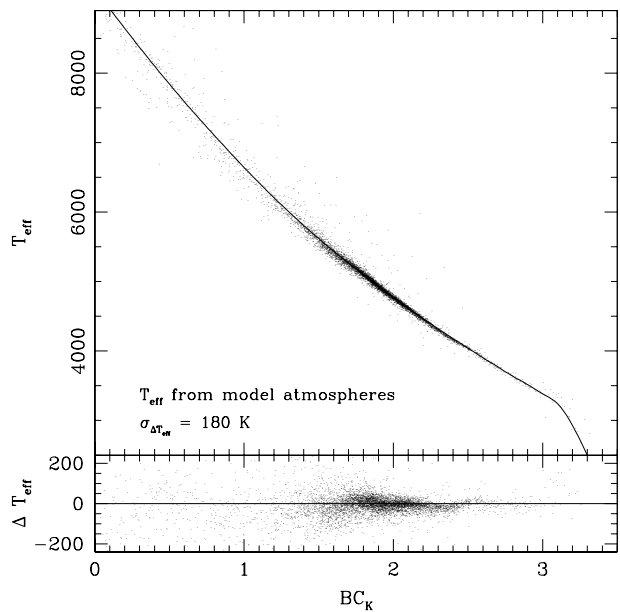}
\caption{
Theoretical T$_{eff}$ derived from averaging the temperatures obtained 
by the color and BC$_K$ as a function of the latter (upper panel). 
The continuous line is our best fit. 
The dispersion (on average 180 K) 
around the fiducial line is also plotted (lower panel). 
}  
\end{figure}

In Fig.7 the theoretical T$_{eff}$ derived from averaging the 
temperatures obtained 
by the color and BC$_K$ as a function of the latter are 
plotted together with the dispersion (on average 180 K) 
around the fiducial line and in Table 3 we listed the 
values.

\subsubsection{Comparison between the two scales}

Comparing the empirical and theoretical temperature scales discussed 
in the previous sections, we infer the following behaviours:
\begin{itemize}
\item in the range 4000 -- 7500 K, a general good agreement has been found, 
the theoretical scale being systematically warmer by about 50 K;
\item between 3200 and 4000 K, this difference increases up to about 100 K;
\item below 3200 K, the two scales diverge, the theoretical one becoming 
progressively cooler, up to 800 K. 
\end{itemize}

The behaviour in the coolest temperature range can be explained as 
a gravity effect, as already mentioned in Sect.3.2.2, in the sense that 
the interpolation in gravity may be quite uncertain.

\section{Discussion}

\subsection{A few more tests}

\begin{figure} 
\epsffile{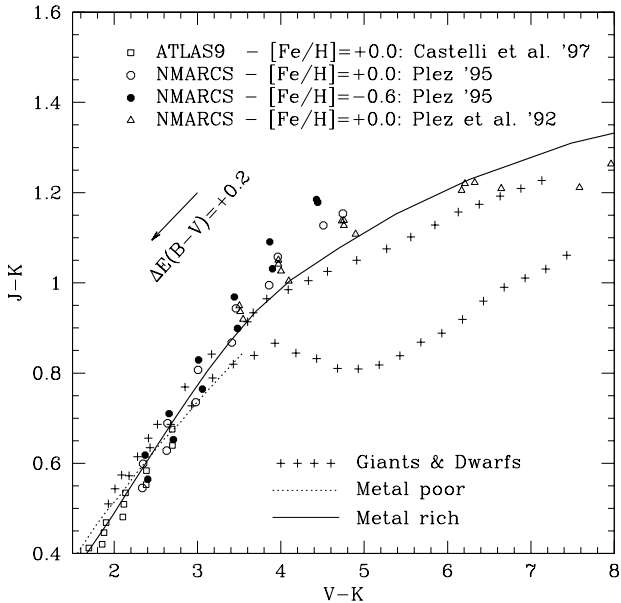}
\caption{
(J--K) {\it vs} (V--K) color -- color diagram for the mean loci 
of the metal rich (continuous line) and metal poor (dashed line) 
observed stars and different model atmospheres. 
ATLAS9 models by Castelli et al. (1997) are without overshooting.
The mean sequences of a sample of field giants and dwarfs by Bessell 
\& Brett (1988) and the reddening vector are plotted as well. 
}  
\end{figure}

In order to check the consistency between our observed database and the 
adopted model atmospheres, both used in the calibrations of the 
temperature scale (see Sect.3.2.2), we constructed a (J--K) {\it vs} 
(V--K) color -- color diagram in the ESO photometric standard system 
for all the stars in our database (see Fig.8).
The mean loci of the observed metal poor and metal rich stars,  
different models, and the mean sequences of a sample of field 
giants and dwarfs as reported in Table 2 and 3 of Bessell \& Brett (1988) 
are plotted as well.
The observed and synthetic distributions well overlap one to each other 
and also overlap the field giant sequence, indicating that the two 
photometric samples are fully consistent. 

Another interesting test we performed is to apply our BC$_K$ and 
T$_{eff}$ relations to the widely used  isochrones computed by Bergbusch 
\& VandenBerg (1992), the only ones presently published in the IR--planes. 
Bell (1992, and reference therein) has shown how these theoretical 
isochrones 
transformed into the observational plane using the Bell \& Gustafsson 
(1989, hereafter BG89) model atmospheres hardly reproduce the sequence of 
cool stars observed by Glass (1974a,b), being the former too red by 
$\approx $0.1 mag along the RGB.       

The results of our test are plotted in Fig.9: the sequences obtained using 
our new calibrations well overlap those computed by using BG89 and
corrected as suggested by Bell (1992) along the RGB (continuous lines), 
while they overlap the uncorrected ones in the turn--off region 
(dashed lines). Hence, it is clear that the discrepancy found  by 
Bell (1992) along the RGB was not due to the theoretical isochrones 
but should rather be a consequence of the adopted transformations into 
the IR observational plane. 

\begin{figure} 
\epsffile{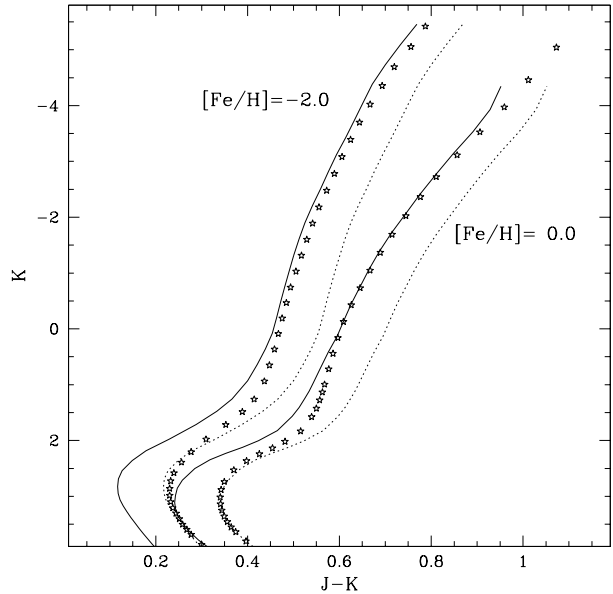}
\caption{
K {\it vs} (J--K) diagram 
of the isochrones by Bergbusch \& VandenBerg (1992) transformed into the 
observational plane at two different metallicities: {\it dotted lines} are 
the relations inferred using BG89 transformations, {\it continuous lines}
represent those corrected as suggested by Bell (1992) to reproduce the 
observed RGB distribution,  and {\it stars}  are those obtained
using our BC$_K$ and T$_{eff}$ transformations. 
}
\end{figure}

\subsection{Comparison with published BC scales}

Our BCs, as reported in Table 3, 
in both the metal poor and metal rich regimes,
were then compared with published scales.

In the BC$_K$ {\it vs} (V--K) plane, we considered  
the relations presented by 
Johnson (1966), Lee (1970), Ridgway et al. (1980), DBR87 and 
BG89, as tabulated by 
DB93 and scaled by  0.01 mag to take into account 
the different BC$_{\odot}$ adopted.
Moreover, we considered  those obtained by FPC81, Bessell \& Wood (1984), 
Blackwell \& Petford (1991) and Alonso et al. (1995). And, finally, 
the BC$_K$ of the model (BCP97), as computed to derive T$_{eff}$ (see 
Sect.3.2.2), were also included. 

\begin{figure} 
\epsffile{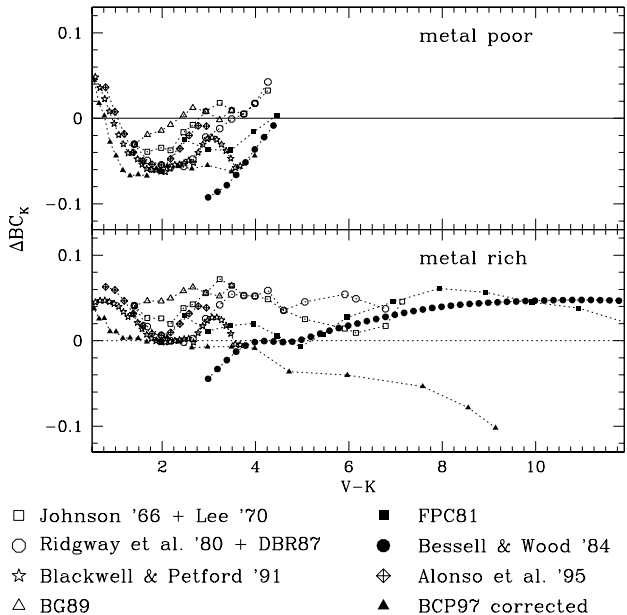}
\caption{
Comparison between our BC$_K$ {\it vs} (V--K) scale as reported in Table 3 
and those obtained by other authors for metal rich (lower panel) and 
metal poor (upper panel) stars (see keys in the plot and Sect.4.2). 
}
\end{figure}

\begin{figure} 
\epsffile{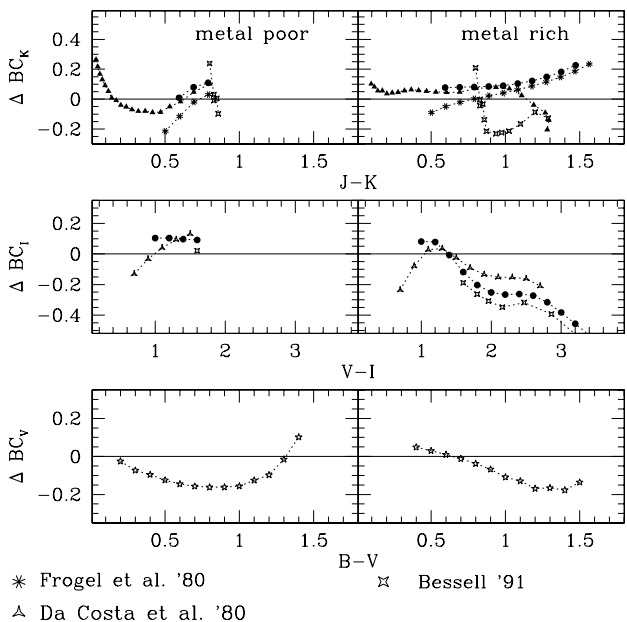}
\caption{
Comparison between our 
BC$_K$ {\it vs} (J--K), BC$_I$ {\it vs} (V--I) and  BC$_V$ {\it vs} (B--V) 
scales and those by other authors for metal poor (left panels) and 
metal rich (right panels) stars. Keys are as in Fig.9, otherwise indicated.
}
\end{figure}

The residuals (our -- others) {\it vs} the (V--K) color are plotted in 
Fig.10 for metal rich 
(lower panel) and metal poor (upper panel) stars. As can be seen,
all the scales agree one to each other within 0.1 mag. 
Interesting enough is the opposite behaviour of $\Delta$BC$_K$ for metal 
rich and metal poor stars. While our BC$_K$ are systematically larger 
than the other scales in metal rich stars, they are systematically lower 
in metal poor ones for a given (V--K).
This may reflect the fact that all the other scales average the 
information of the two metallicity regimes.

Concerning BC$_K$ as a function of (J--K), we compared our scale with 
those presented by Frogel, Persson \& Cohen (1980), Bessell \& Wood (1984), 
Bessell (1991) and with the BC$_K$ of the model (BCP97).

The relation in the BC$_I$ {\it vs} (V--I) plane was 
compared with those listed by Bessell \& Wood (1984), Da Costa \& Armandroff 
(1990) and  Bessell (1991).

Finally, the BC$_V$ {\it vs} (B--V) relation 
was compared with that obtained
by Blackwell \& Petford (1991).

The residuals drawn from each individual comparison  
{\it vs} the adopted colors
are plotted  in Fig.11 for both metal poor (left panels) and metal rich 
(right panels) stars. The average spread among different scales is 
larger in these planes than that inferred in the BC$_K$ {\it vs} 
(V--K) plane.

\subsection{The IRFM}                          

Using our BC$_K$ and empirical T$_{eff}$ we can also provide an 
independent, empirical calibration of the R parameter. 
This parameter was computed  using the same ZP of our bolometric scale and 
the K band flux calibration of Table 2 (columns \#6 and \#7). 
The results are listed in Table 4.

This empirical scale has been compared with the relation obtained 
interpolating the theoretical Blackwell \& Lynas--Gray (1994) grid, 
which tabulated log~R as a function of T$_{eff}$ and log~g. 
The interpolation in gravity 
was performed using the VandenBerg (1996) isochrones 

In Fig.12 we plot the residual T$_{eff}$ between the two relations as 
a function of our empirical log~R and T$_{eff}$.
For log~R$>$1 there is a good agreement ($\Delta$T$_{eff}<\pm$ --50 K)  
among the two relations.
For log~R$<$1 the empirical T$_{eff}$ seem to be systematically hotter 
than those of the Blackwell \& Lynas--Gray (1994) models.

\begin{figure} 
\vspace{-4.5cm}
\epsffile{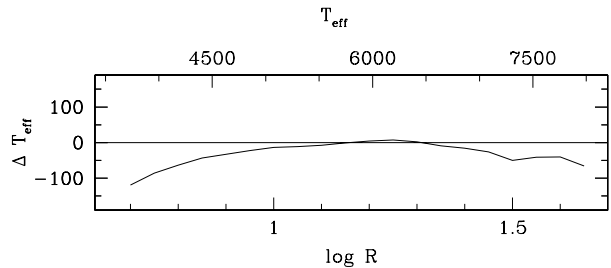}
\caption{
$\Delta$T$_{eff}$ {\it vs} log~R and empirical T$_{eff}$ between 
our empirical scales and the theoretical ones by Blackwell \& Lynas--Gray 
(1994). 
}
\end{figure}

\begin{table}
 \centering
 \begin{minipage}{80mm} 
  \caption{ R--parameter of the IRFM as a function of T$_{eff}$ and BC$_K$.}
  \begin{tabular}{@{}lccccc@{}}

T$_{eff}$ &BC$_K$  &log~R$^{\dag }$  &  T$_{eff}$ &BC$_K$  i
&log~R$^{\dag }$ \\ 
  3250 & 3.10 & 0.5504  &   5750 & 1.41 & 1.2235 \\
  3500 & 2.82 & 0.6613  &   6000 & 1.28 & 1.2764 \\
  3750 & 2.65 & 0.7303  &   6250 & 1.15 & 1.3278 \\
  4000 & 2.47 & 0.7993  &   6500 & 1.03 & 1.3781 \\
  4250 & 2.30 & 0.8676  &   6750 & 0.90 & 1.4275 \\
  4500 & 2.14 & 0.9327  &   7000 & 0.78 & 1.4757 \\
  4750 & 1.98 & 0.9953  &   7250 & 0.66 & 1.5232 \\
  5000 & 1.83 & 1.0555  &   7500 & 0.55 & 1.5698 \\
  5250 & 1.69 & 1.1134  &   7750 & 0.43 & 1.6158 \\
  5500 & 1.55 & 1.1692  &        &      &        \\
\end{tabular}
\vspace {0.2cm}
{\footnotesize
\par\noindent
{\bf $^{\dag }$~}logarithmic value of R = ${F_{bol}\over{F_{2.2 \mu m}}}$ 
in units of $\mu $m.}    
\end{minipage}
\end{table}

\section{Conclusions}

The main results obtained from our analysis can be summarized as follow:
\begin{itemize}
\item By exploiting the use of  a large photometric database of Pop II 
stars in GGCs we derived  new relations for:
BC$_K$ {\it vs} (V--K) and (J--K),
BC$_V$ {\it vs} (B--V), (V--I) and (V--J), 
to infer empirical bolometric corrections from observed colors. 
\item By making use of both an empirical relation and model atmospheres 
we calibrated two different scales to infer reliable stellar effective 
temperatures in the range 3000 -- 7500 K.
\item We also calibrated an empirical relation between the R 
parameter of the IRFM and the stellar temperature. 
\end{itemize}
All these relations are summarized in Table 3, where 
for a given BC$_K$ the corresponding colors and BC$_V$
in two different metallicity regimes and T$_{eff}$ can be read.  
These relations should be the most suitable to 
calibrate the red stellar sequences in the color--magnitude diagrams, 
like the RGB and AGB in old GGCs. 

\section*{Acknowledgments}
We wish to thank all the people who kindly made available to us 
photometric data, model atmospheres, and isochrones. We warmly thank Mike
Bessell for his very careful reading of the paper, 
Fiorella Castelli, Giampaolo Di Benedetto, Maria Lucia Malagnini 
and Bob Kurucz for helpful discussions. A special thank to the ADC Service to 
provide the Morel \& Magnenat (1978) Catalogue. Paolo Montegriffo was supported 
by a 1996-grant of the {\it Fondazione del Monte, Rolo Banca 1473}.
For their financial support, we also thank the 
{\it Ministero della Universit\`a e
della Ricerca Tecnologica} (MURST) and the {\it Agenzia Spaziale
Italiana} (ASI).

%%%%%%%%%%%%%%%%%%%%%%%%%%%%%%%%%%%%%%%%%%%%%%%%%%%%%%%%%%%%%%%%%%%%%%%%%%%%
\end{document}